\documentclass[10pt]{iopart}
\usepackage[T1]{fontenc}
\usepackage[latin9]{inputenc}
\usepackage{geometry}
\usepackage{cite}
\usepackage{graphicx}
\usepackage{color}
\makeatletter
\usepackage{babel}

\makeatother

\begin{document}

\title{Nanomechanical Design Strategy for Single-Mode Optomechanical Measurement}
\author{G. La Gala, J. P. Mathew, P. Neveu and E. Verhagen}
\address{Center for Nanophotonics, AMOLF, Science Park 104, 1098 XG Amsterdam, The Netherlands}

\begin{abstract}
The motion of a mechanical resonator is intrinsically decomposed over a collection of normal modes of vibration. When the resonator is used as a sensor, its multimode nature often deteriorates or limits its performance and sensitivity. This challenge is frequently encountered in state-of-the-art optomechanical sensing platforms. We present a mechanical design strategy that ensures that optomechanical measurements can retrieve information on a single mechanical degree of freedom, and implement it in a sliced photonic crystal nanobeam resonator. A spectral design approach is used to make mechanical symmetries robust against practical disorder. The effectiveness of the method is evaluated by deriving a relevant figure of merit for continuous and pulsed measurement application scenarios. The method can be employed in any mechanical design that presents unwanted spurious mechanical modes. In the nanobeam platform, we experimentally show an increase of the signal to noise ratio of the mode of interest over the first spurious mode by four orders of magnitudes.
\end{abstract}
\maketitle
\ioptwocol

\section{Introduction}
In optomechanical systems, co-localizing light and mechanical oscillations enables ultraprecise sensing and optical control over mechanical motion \cite{Aspelmeyer2014CavityOptomechanics,Degen2017}. A variety of experiments explore mechanical resonators of different masses and sizes. The properties of these systems and their ability to interact with external systems allow conceptually different applications ranging from gravitational wave sensing \cite{Abbott2016ObservationMerger}, compact sensors of various external stimuli \cite{Krause2012,Simonsen2019,Ghosh2020,Westerveld2021,Bemani2021}, and stochastic heat engines \cite{Maslennikov2019,Serra-Garcia2016,Mari2015,Fernandez-Alcazar2021}, to quantum mechanical entanglement \cite{Riedinger2018,Kotler2021,Ockeloen-korppi2021} or photon-phonon transducers in quantum networks  \cite{Zhang2003Quantum-stateOscillators,Hill2012,Riedinger2016}. Besides applications, optomechanics allows fundamental studies into the quantum nature of macroscopic mechanical objects \cite{Aspelmeyer2014,Ananyeva2021} using pulsed \cite{Vanner2011PulsedOptomechanics.,Muhonen2019StatePulses,Neveu2021} or continuous \cite{Wieczorek2015,Rodrigues2021,Meng2020,magrini2021real,Tebbenjohanns2021} measurements, and investigation of possible decoherence mechanisms \cite{Pino2018On-chipMicrosphere, Chen2013MacroscopicOptomechanics}. The class of nano- and micro-mechanical systems is particularly relevant in this field of study, thanks to the tremendous efforts towards ultracoherent resonators \cite{Beccari2021,Serra2021,Hoj2021}.

For many optomechanical applications, one typically wishes to work with a single mechanical mode, or at least isolated from others. In a quantum control perspective, other modes affecting the measured signal cause noise on that estimation, and thus limit control. However, a typical mechanical system will intrinsically display a collection of modes of vibration, rather than just a single one. The vibrational modes supported by a structure can be complex, depending on the structure's geometry, and can be fundamentally understood in the context of the theory of elasticity to form a collection of normal modes. This multimodeness can be a limitation in feedback cooling experiments where the signal of an optical mode whose phase is sensitive to mechanical motion is used to produce a negative feedback on the oscillator to damp its motion \cite{Wilson2015Measurement-basedRate,Rossi2018Measurement-basedMotion,Guo2019, Lai2020,Bemani2021,Ananyeva2021}. Likewise, besides specific advantageous situations where the ratio of mechanical modes' frequencies is integer \cite{Neveu2021}, it limits the ability of a pulsed position measurement to estimate one quadrature of a single mechanical resonator while evading backaction, as any instantaneous measurement probes all modes at once \cite{Vanner2011PulsedOptomechanics.,Muhonen2019StatePulses}.

Optomechanical design can be used to isolate the optical sensitivity to a single favorable mechanical mode. In particular, the use of structural symmetries can decrease the sensitivity to spurious modes in various cases such as nanobeam cavities \cite{Leijssen2015StrongNanobeam,Safavi-Naeini2013SqueezedResonator,Eichenfield2009ACavity,Piergentili2018,Gartner2018}. There, unwanted mechanical modes do not couple to the optical cavity mode as their displacement does not affect the effective cavity length by symmetry reasons. Then, any fabrication imperfection that breaks symmetry limits the effectiveness of this method in practice, as it can induce symmetry breaking in the mechanical modes of interest, such that unwanted mechanical modes couple to the cavity. Here, we present and experimentally implement a strategy to mitigate the effect of such unwanted mechanical modes, `purifying' the optomechanical spectral response of a structure around a single mechanical degree of freedom. In section \ref{sec:fom}, we examine in detail the importance of noise contributions due to the presence of spurious modes, quantifying these by defining a suitable figure of merit in a continuous measurement. We then argue how `spectral design' of the system can optimize this figure of merit even in the presence of perturbations due to random fabrication imperfections in section \ref{sec:design}. We implement in section \ref{sec:results} and \ref{sec:pulsed} this design strategy in the specific example of a sliced photonic crystal nanobeam cavity, and show a reduction of the noise induced by other mechanical modes by four (one) orders of magnitudes in a practical continuous (pulsed) measurement. Finally, in section \ref{sec:coupledmode}, we discuss how the strategy can be interpreted in a model of two coupled modes, and demonstrate how it can be implemented also with the assistance of the optical spring effect. 

\section{Noise in optomechanical measurements induced by spurious modes}
\label{sec:fom}

We first examine the scenario in which a continuous reflectance measurement is performed on an optomechanical system having a pair of mechanical modes with resonance frequencies $\Omega_1$ and $\Omega_2$ that are nearly degenerate, and that both couples weakly to the same meter. We treat the presence of the spurious mode as extra noise sources. We characterize the contribution of the excess imprecision noise due to the presence of the mode at $\Omega_1$ in a measurement of that at $\Omega_2$ (the mode of interest) in the case of continuous measurement (See section \ref{sec:pulsed} for the case of pulsed measurement). In the presence of two independent mechanical modes, the cavity frequency $\omega_c$ can be written as
\begin{equation}
\omega_c = \bar\omega_c+ G_1 x_1+G_2 x_2,
\label{w_c}
\end{equation}
where $\bar\omega_c$ is the optical mode frequency at rest position of the resonators, $x_{i}$ the displacement and $G_{i}=\partial\omega_c/\partial x_i $ the optical frequency shift per unit of displacement of mode $i$ with mass $m_i$. Displacement of each mode can be expressed in unit of its zero-point motion amplitude $x_{\mathrm{zpf},i}=\sqrt{\hbar/2m_i\Omega_i}$, with corresponding detuning $g_i=G_i x_{\mathrm{zpf},i}$ the vacuum optomechanical coupling rate.

\begin{figure}
\centering
\includegraphics[width=8 cm]{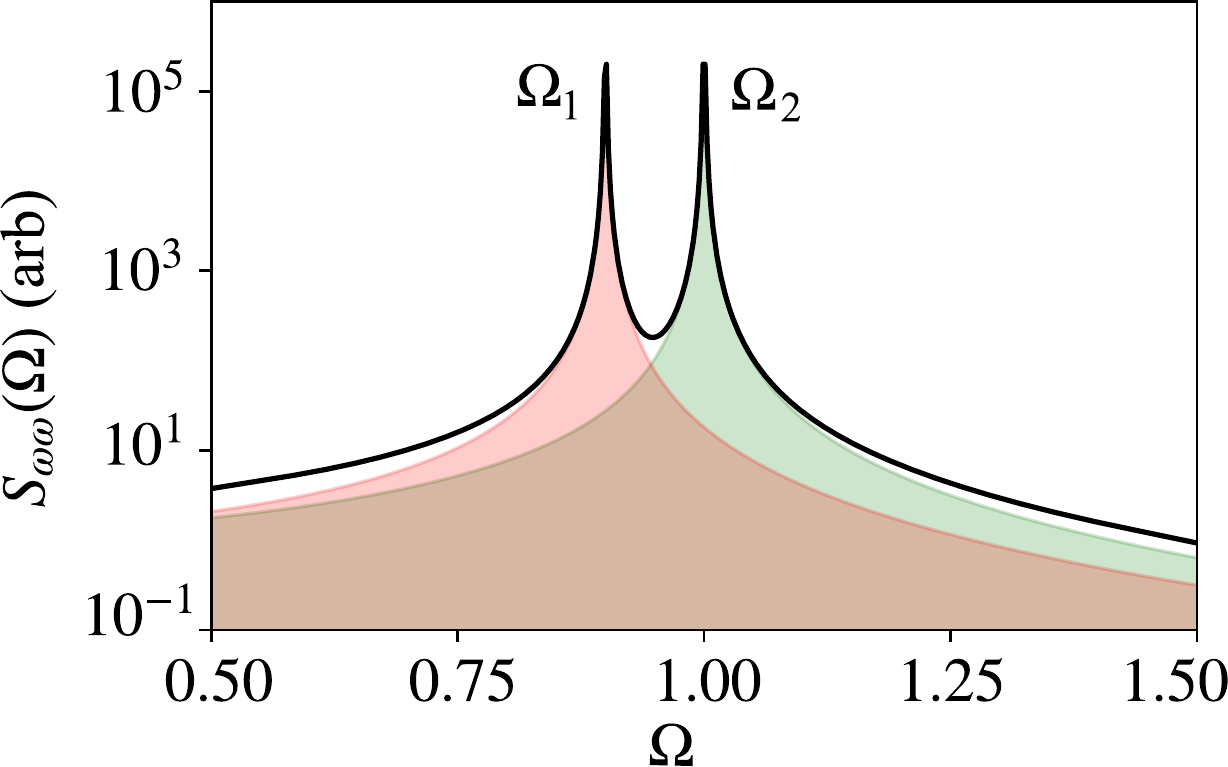}
\caption{Illustration of typical frequency fluctuation spectra due to two mechanical resonances at frequencies $\Omega_1$ and $\Omega_2$. The noise spectral density originating from the mode of interest 2 (spurious mode 1) is filled in green (red). The wing of the Lorentzian of mode 1 is directly linked to the added imprecision noise at the resonance frequency of mode 2. To characterize the imprecision added to a measurement of mode 2 due to the presence of mode 1, we look at the value of the fluctuations of mode 1 at the frequency $\Omega_2$.}
\label{drawing PSD}
\end{figure}

In a continuous measurement, an optomechanical device is probed by a light field with constant intensity over a time span that is much longer than a single mechanical period $2\pi/\Omega_2$. The mechanical fluctuations that act as a parametric modulation of the optical resonance frequency (equation\,\ref{w_c}) can be studied in the frequency domain as fluctuation spectra  (figure\,\ref{drawing PSD}). As we derived in \ref{ContinuousFOM}, one can quantify the added imprecision incurred by the presence of the unwanted mechanical mode detuned by $\Delta\Omega = \Omega_2 - \Omega_1\ll\Omega_{1,2}$, by the figure of merit

\begin{equation}\label{fig_merit_CW}
\frac{S_{\omega\omega}^\mathrm{th,1}}{S_{\omega\omega}^\mathrm{zpf,2}}=2\frac{k_B T}{\hbar} \Gamma_1\Gamma_2 \frac{g_{1}^2}{g_{2}^2} \frac{\Omega_1}{\Delta\Omega^2(\Delta\Omega+2\Omega_1)^2} 
,\end{equation}
where, on the left hand side, $S_{\omega\omega}^\mathrm{th,1}$ is the frequency power spectral density of the detrimental mode, quantifying the impact of position fluctuations of mode 1 at the frequency of mode 2. This noise amount is normalized to
\begin{equation}
S_{\omega\omega}^\mathrm{\mathrm{zpf}}=g_2^2 S_{x_2x_2}^\mathrm{\mathrm{zpf}}=g_2^2\frac{2 x_\mathrm{\mathrm{zpf},2}^2}{\Gamma_2},
\end{equation}
that describes the spectral density of the zero point motion of the mode of interest. On the right hand side of equation\,\ref{fig_merit_CW}, $k_BT$ is the thermal energy and $\Gamma_i$ the damping rate of mechanical mode $i$ \cite{Note}. 
We see that the imprecision due to the unwanted mode scales with temperature, as the spurious mode population increases with it. It also scales with the ratio between the photon-phonon coupling rates $g_1/g_2$, and with the linewidths of both modes, as it scales effectively with the overlap of the two modes' response spectra. 

A key feature of equation \ref{fig_merit_CW} is the inverse dependence in the spectral proximity of the modes' frequencies $\Delta\Omega$, leading to dramatic detrimental noises when nearly degenerate mechanical modes both couple to the optical cavity. Indeed, in various systems, such as pairs of vibrating beams or membranes, modes of opposite symmetry are close to degeneracy by nature. Even if symmetry can ensure that, in principle, $g_1$ is designed to be zero such that it does not affect the measurement of mode 2, imperfections can break symmetry, leading to an unwanted finite $g_1$. In the next section we show how it is possible to define a strategy for the design of a more robust solution that relies on releasing the degeneracy of the normal modes, in order to enforce mode symmetry also in the presence of disorder. 

\section{Robust mode symmetries in optomechanical systems}

Understanding the influence of disorder to symmetry breaking can be gained through perturbation theory. Here we will consider how a pair of orthogonal modes (i.e. \textit{normal modes}) $\psi_1^{(0)}$ and  $\psi_2^{(0)} $ that are near in energy but not exactly degenerate are affected by disorder. We follow the formalism of Cotrufo et al. \cite{Cotrufo2017ControlPerturbation}. The disorder can be described by a potential $\hat{V}(r,t)$, which modifies the eigenvalues of the system. In nondegenerate perturbation theory, one can calculate the new energy eigenvalues by 
\begin{eqnarray}
E_1^{'} &= E^{(0)}_1 + \left< \psi_1^{(0)} \left|\hat{V}\right| \psi_1^{(0)} \right>,\\
E_2^{'} &= E^{(0)}_2 + \left< \psi_2^{(0)} \left|\hat{V}\right| \psi_2^{(0)} \right>,
\label{eq pert energies}
\end{eqnarray}
where $E^{(0)}_i$ are the eigenvalues for the bare system, in the absence of perturbation. The disorder potential couples the modes and adds the extra term with respect to the initial \textit{bare} eigenvalues $E^{(0)}_1$ and $E^{(0)}_2$ in equation\,\ref{eq pert energies}.

In that case, the potential has off-diagonal elements and modify the eigenfunctions as well, giving rise to a different spatial configuration of the modes. This can be written as
\begin{eqnarray}
\left|\psi_1^{'}\right> &= \left|\psi_1^{(0)}\right> + \frac{\left< \psi_2^{(0)} \left|\hat{V}\right| \psi_1^{(0)} \right>}{E^{(0)}_1 - E^{(0)}_2} \left|\psi_2^{(0)}\right>,\label{eq pert eigenveca}\\
\left|\psi_2^{'}\right> &= \left|\psi_2^{(0)}\right> + \frac{\left< \psi_1^{(0)} \left|\hat{V}\right| \psi_2^{(0)} \right>}{E^{(0)}_2 - E^{(0)}_1} \left|\psi_1^{(0)}\right>.
\label{eq pert eigenvec}
\end{eqnarray}
Equations \ref{eq pert energies} and \ref{eq pert eigenvec} show that in the presence of a disorder potential both the energy eigenvalues (i.e. the frequencies) and the eigenvectors (i.e. the spatial shape) are modified. In particular, equation \ref{eq pert eigenvec} shows that the correction in the spatial configuration is such that it has the symmetry of the orthogonal mode. Indeed, if modes $\left|\psi_1^{(0)}\right>$ and $\left|\psi_2^{(0)}\right>$ are anti-symmetric and symmetric, respectively, any asymmetrically distributed disorder $\hat{V}$  will make the numerators $\left< \psi_{1,2}^{(0)} \left|\hat{V}\right| \psi_{2,1}^{(0)} \right>$ in equation \ref{eq pert eigenvec} nonzero. The consequence of this perturbation is to progressively mix the shape of the two modes as its importance increases, destroying the initial symmetries of the individual eigenmodes. For example, an initially perfectly symmetric mode may lose that quality as an asymmetric disorder mixes it with a mode of orthogonal symmetry. In the next section we will show a practical observation of this effect in a specific geometry of interest.

In practical experiments, one cannot fully eliminate perturbation incurred in the fabrication process. By inspecting equation\,\ref{eq pert eigenvec}, we recognize a strategy to reduce the \emph{impact} of disorder, as the magnitude of the symmetry breaking is inversely proportional to the energy difference $E_1^{(0)}-E_2^{(0)}$ of the modes of the unperturbed system. Thus, if one intentionally shifts the frequency of the unwanted mode further away from that of the mode of interest, each corrected eigenvector keeps the shape of the unperturbed system solution. Such a design principle thus forces the system to retain the symmetries also in presence of disorder. Controlling mode spectra through geometry is in other words a practical way to render symmetries robust against disorder.
\begin{figure}
\centering
\includegraphics[width=\columnwidth]{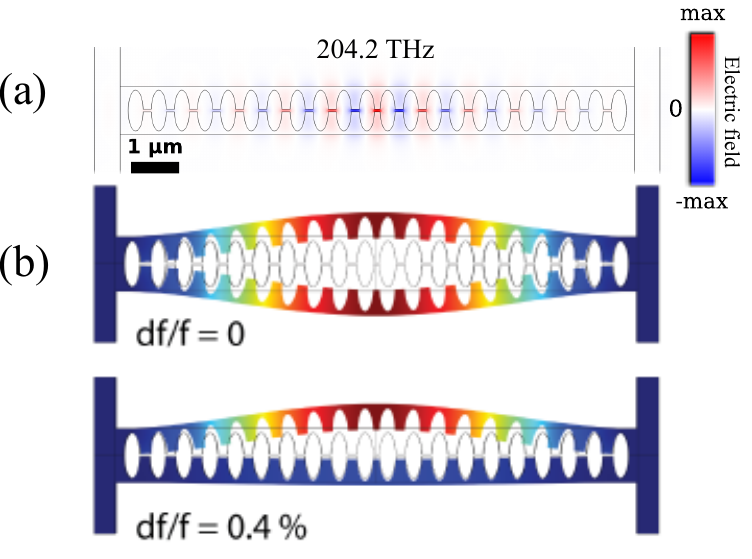}
\caption{Schematics of the sliced silicon nanobeam structure. (a) An optical mode is hosted in a defect in the middle of the structure. (b) When the two halves of the nanobeam are perfect, they both halves have equal amplitude in the symmetric mode (displayed, top). The two halves may however have different mechanical eigenmodes frequency because of slight fabrication disorder. As a result, the two normal modes are neither symmetric nor anti-symmetric, instead predominantly localized in the individual beams (one example shown at the bottom). Here $\mathrm d f$ denotes the frequency difference between the individual beams caused by disorder.}
\label{fig:disorder}
\end{figure}
 
\section{Implementation of mechanical design strategy} \label{sec:design}
The sliced photonic crystal nanobeam that we study consists of two beams that are free to oscillate and share common clamps  (figure \ref{fig:disorder}.a) \cite{Leijssen2017NonlinearFluctuations}. 
Each beam exhibits flexural modes of vibration whose frequency and amplitude are determined by the length and mass density of the beam. In a perfect structure the two beams would be the exact copy of each other, oscillating with the same frequency and amplitude. This kind of structure would present perfect normal modes of vibration: a common mode in which the beams move in plane exactly in phase and a differential mode in which the beams move in anti-phase  (figure \ref{fig:disorder}.b).

From the perspective of optomechanics this idealized case is optimal since only the differential mode couples to the light field.
In actual devices, fabrication imperfections result in a different mass or strain profile of the two beams. In turn this leads to different oscillation frequency (within 3\% in practice) and amplitude for each beam \cite{Leijssen2015StrongNanobeam}. The intrinsic coupling between the two beams through strain in the clamping supports is weak, i.e. the magnitude $|E_1^{(0)}-E_2^{(0)}|$ of the denominators in equations \ref{eq pert eigenveca} and \ref{eq pert eigenvec} is small, such that the breaking of this symmetry forces the hybridization of the modes of vibration of the individual beams to nearly vanish. The result is that each beam oscillates independently from the other  (figure \ref{fig:disorder}.c) and the eigenmodes are neither symmetric nor anti-symmetric. This implies that the sliced nanobeam typically presents two modes that strongly couple to the cavity field and that are nearly degenerate.
\begin{figure}
\centering
\includegraphics[width=\columnwidth]{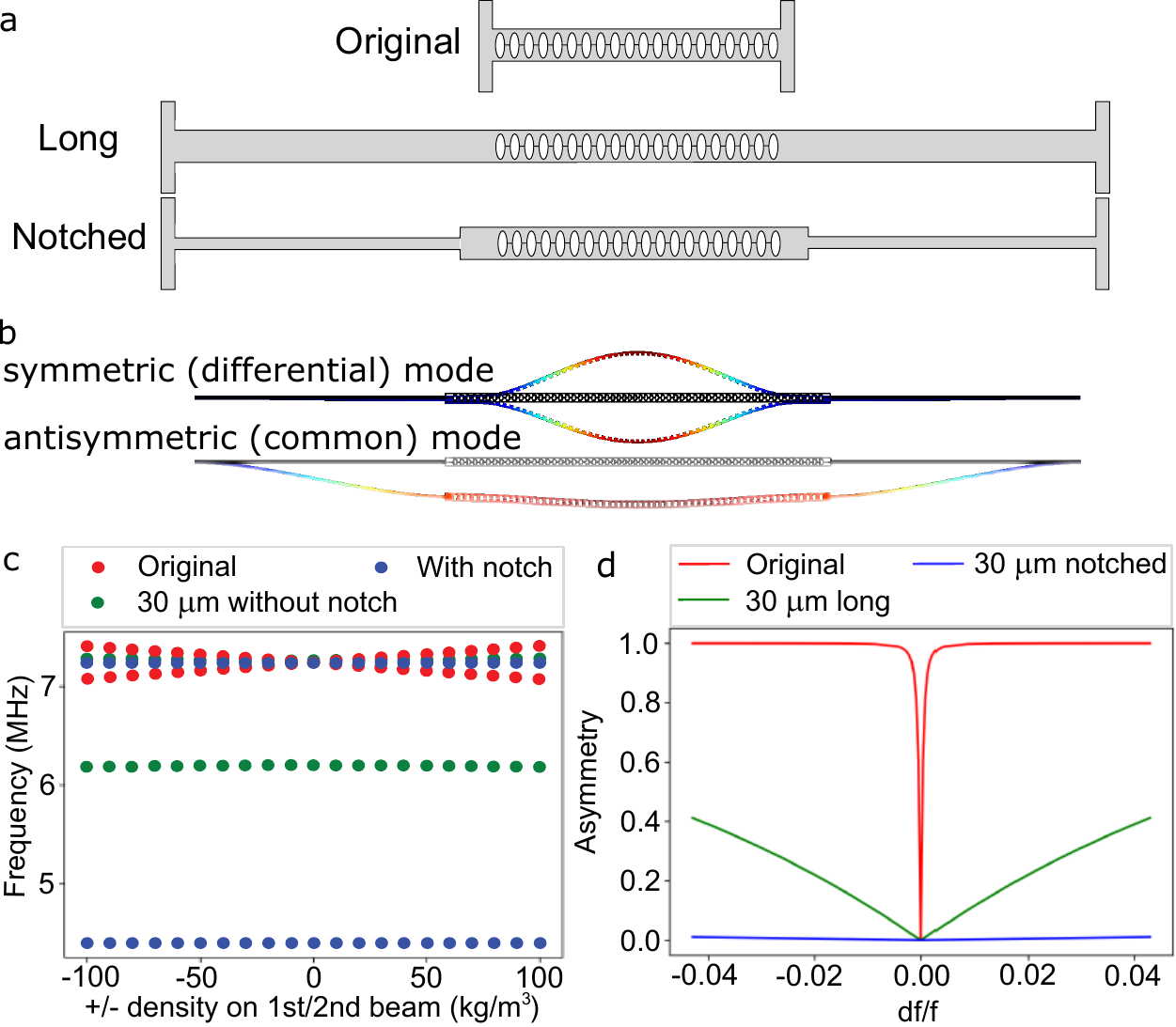}
\caption{(a) Types of geometries that were used in the simulations. (b) Simulated differential and common modes for the notched geometry. Large frequency splitting is induced in the long and notched geometries, because of their motional mass difference.  (c)  Mechanical frequencies, showing the frequency splitting for different designs.  The lower frequency in the long and notched geometry corresponds to the common mode. (d) Asymmetry is much more resilient to disorder in the notched geometry.}
\label{fig:geometry}
\end{figure}
Our aim is to find a nanobeam geometry that has a differential mode that is immune to disorder.

We quantify the asymmetry of this mode as
\begin{equation}
A=1-\left(\frac{x_\mathrm{up}}{x_\mathrm{down}}\right)^2
,\end{equation}
where $x_\mathrm{up}$ and $x_\mathrm{down}$ are the amplitudes of oscillation of the individual beams in the center of each beam. We choose $x_\mathrm{down}$ to be always the larger so that $0<A<1$. The case of $A=0$ corresponds to a perfect symmetric differential mode.

We quantify the strength of the symmetry-breaking perturbation (i.e., the strength of
the disorder), by a parameter $\mathrm d f$  with units of frequency. In the finite-element simulation, the perturbation
is introduced by varying the mass density of one beam half, leading to a frequency splitting $\mathrm d f$ of the two halves' fundamental flexural modes. In figure\,\ref{fig:geometry}, we investigate numerically how the asymmetry varies with disorder, for structures in which we increase the length of the nanobeam outside of the photonic crystal region and the case in which we create a notched extension also outside the photonic crystal. Increasing the length of the beam extension, at fixed length of the sliced beam, lowers the frequency of the common mode, while keeping that of the differential hardly changed. The notched design decreases the in-plane flexural stiffness of the beam, thus further lowering the frequency of the common mode away from that of the differential mode.

Figure \ref{fig:geometry}.d shows that the notched geometry gives additional protection of the differential mode against disorder by pushing the common mode to lower frequencies while maintaining the differential mode frequency close to the original. The original structure is seen to quickly become fully asymmetric for small values of frequency asymmetry $\mathrm d f / f = 0.001$. In contrast, the notched structure shows good stability against disorder even at a disorder strength $\mathrm d f / f \simeq0.04$. From experiments we know the disorder level of the original structure is about $\mathrm d f/f = 0.03$, leading to complete lifting of mode degeneracy. The alternative designs, in particular the notched structure, feature a much greater resilience to disorder as the frequencies of the differential and common modes in the unperturbed structure are pushed much further apart. It should be noted that the notched structure becomes floppy for any higher frequency common flexural modes, and lowering its frequency bring it closer to that of the fundamental differential mode. This puts a practical constraint on how far the fundamental common mode can be pushed away.

\section{Experimental Results}
\label{sec:results}

\begin{figure}
\centering
\includegraphics[width=0.8\columnwidth]{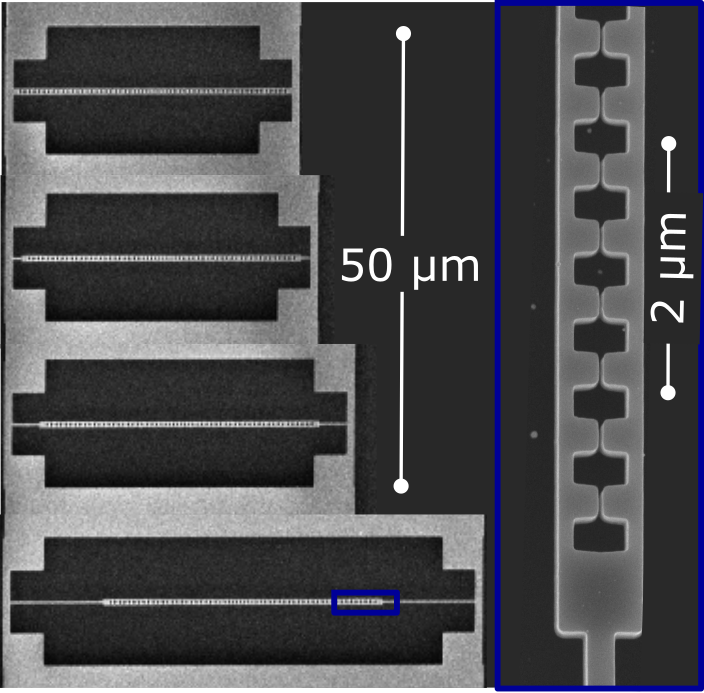}
\caption{Fabricated devices that share the same photonic crystal in the middle (top). From left to right the length of the extension on both sides of the photonic crystal is of 0,1,3 and 10 $\mu$m. The bottom picture is a zoom in on one of the devices.}
\label{sem_pic}
\end{figure}

We fabricate sliced photonic crystal nanobeams in silicon that have in common the same photonic crystal design in the middle, but vary in their connection to the outside, as per the design strategy of figure \ref{fig:geometry}. The structures are fabricated in the 220\,nm thick silicon layer of a silicon-on-insulator substrate using electron beam lithography and reactive ion etching with HBr/O2 plasma \cite{Leijssen2015StrongNanobeam}. The devices are released by etching the 3\,$\mu$m thick buried oxide layer with hydrofluoric acid followed by critical point drying. Figure \ref{sem_pic} reports scanning electron micrographs of the structures, fabricated on a single chip. We vary the length of the notched extension that is symmetrically placed on both sides of the photonic crystal region. The optomechanical coupling rate of those structures is typically about $g_0\simeq2\pi\times30\,$MHz \cite{Leijssen2017NonlinearFluctuations}.

\begin{figure}
\centering
\includegraphics[width=\columnwidth]{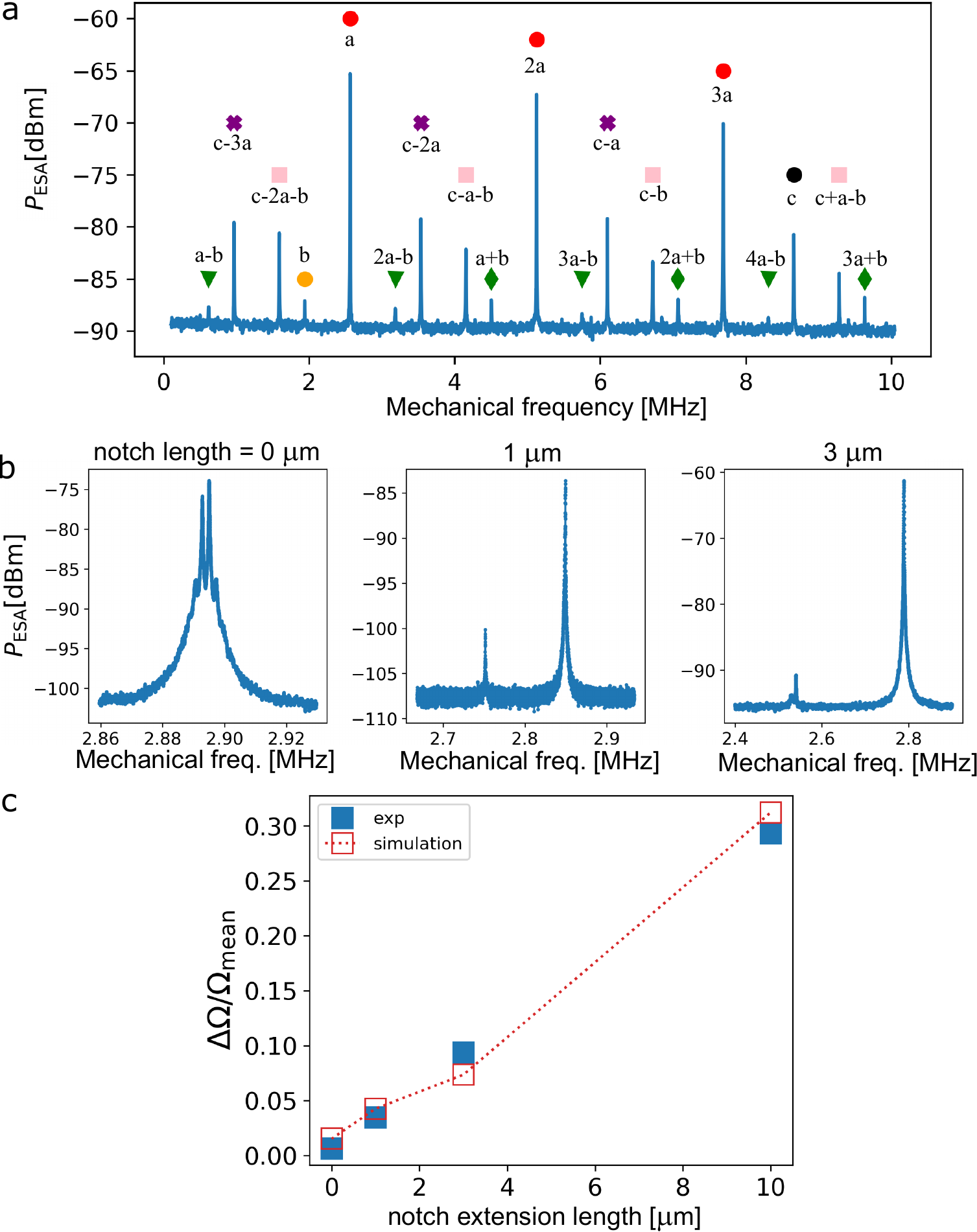}
\caption{(a) Measured fluctuation spectrum for a structure with 10 $\mu m$ notch extension. The spectrum has a bandwidth of 10 MHz and shows multiple mechanical peaks. Three distinct mechanical modes are detected within this range ($a=2.56$ MHz, $b=1.94$ MHz and $c= 8.66$ MHz). Other peaks correspond to mixing of these three by nonlinear transduction. (b) Spectra of four structures with notch length of 0,1,3 and 10 $\mu$m, illustrating the typically reduced magnitude of the spurious mode of longer notch length. (c) Relative frequency difference comparison between numerical simulations and experiment. }
\label{spectra_exp}
\end{figure}

In order to characterize the mechanical performance we perform optical reflection measurements on the devices (see Supp. Mat. of \cite{Muhonen2019StatePulses} for more details on the experimental characterization setup). A focused laser beam that is tuned to resonance with the optical mode is used to excite the cavity and the reflected signal from the cavity is collected by a homodyne interferometer. The signal carries a modulation that is due to the mechanical thermal fluctuations at room temperature. The Fourier transform of the voltage signal generated by the balanced photodetector of the interferometer displays the mechanical fluctuation spectrum, and is measured  with an electronic spectrum analyzer. The signal $P_\mathrm{ESA}$ is then directly proportional to the spectral density $S_{\omega\omega}$. A typical spectrum is shown in figure \ref{spectra_exp}.a. We can recognize a multitude of fluctuation peaks.  Only three of those correspond to distinct mechanical mode frequencies, namely the first-order in-plane flexural modes (differential and common), and the third-order in-plane flexural differential mode. respectively labeled "a", "b", and "c" in Fig.\,\ref{spectra_exp}a. The other peaks are a signature of the large optomechanical coupling that gives rise to nonlinear transduction \cite{Leijssen2017NonlinearFluctuations} and hence to the generation of higher harmonics in the spectrum that occur at sums of integer number of bare mode frequencies. Figure\,\ref{spectra_exp}.b shows spectra with a span limited around modes "a" and "b", for devices of the kind shown in figure\,\ref{sem_pic}. 
By studying the spectra in figure \ref{spectra_exp}.a, it is then possible to characterize the mechanical response around the frequency of interest. The differential flexural mode  (figure\,\ref{spectra_exp}.b) of the sliced nanobeam is the higher frequency mode. Its frequency slightly lowers when the nanobeam notch length becomes longer. This is potentially related to the change of the precise clamping point geometry or of the remaining compressive stress  in the silicon device layer \cite{Leijssen2015StrongNanobeam}. At the same time, as the notch extension becomes longer, the lower frequency mode shifts significantly towards lower frequencies. This shift reflects the increase in the overall length of the device.  We note that the two side-peaks in the leftmost spectrum of figure \ref{spectra_exp}.b are again due to the mixing of the two modes related to nonlinear transduction \cite{Leijssen2017NonlinearFluctuations}. In figure \ref{spectra_exp}.c an overview of the relative mechanical frequencies for different notch lengths is shown, together with results obtained by calculations of the mechanical eigenfrequencies obtained for the same geometries simulated with finite-element software COMSOL. The experimental results we show are averaged over three realizations but statistical variations in disorder can still lead to deviations from trends predicted for fixed disorder. 

\begin{figure*}
\centering
\includegraphics[width=\textwidth]{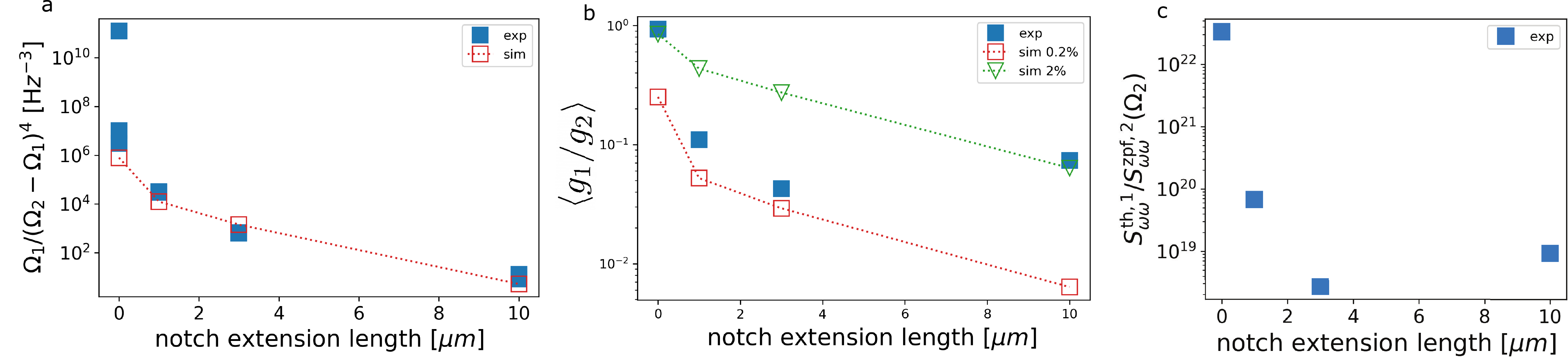}
\caption{Optomechanical purification, according to the figure of merit of equation\,\ref{fig_merit_CW}. (a) Frequency factor present in the figure of merit for continuous measurements. We rescale the frequencies in order to take into account the absolute value mismatch between simulated and measured mechanical resonance frequencies. (b) Ratio between the coupling strength of mode 1 and 2 versus extension length. Two simulation curves are shown: for disorder equal to $\mathrm df/f=0.002$ (empty red squares, dashed line), and for disorder equal to $\mathrm d f/f=0.02$ (empty green triangles, dashed line). (c) The added imprecision noise from mode 1 in a measurement of mode 2 as a function of the beam extension.}
\label{g0s}
\end{figure*}

In order to verify the efficiency of our design we now evaluate the figure of merit (\ref{fig_merit_CW}).
We use the Wiener-Kinchin theorem to relate the 'band power' $BP_i$ (the integral of fluctuations of mode $i$) directly to the thermal fluctuations that populate the mechanical modes
\begin{eqnarray}\label{bp}
BP_i &= \int_{\Omega_i-\delta}^{\Omega_i+\delta} S_{\omega\omega} d\Omega = \left< \delta\omega^2 \right>_{\mathrm{th},i}\nonumber \\
& = 
G_i \left< x_i^2\right>_\mathrm{th} = 2 n_{\mathrm{th},i} g_i^2
\end{eqnarray}
where for the last step we have used the expression for the variance of the mechanical displacement induced by thermal fluctuations $\left<x^2\right>_\mathrm{th}=2x_\mathrm{\mathrm{zpf}}^2 n_\mathrm{th} = 2x_\mathrm{\mathrm{zpf}}^2 k_B T/\hbar\Omega$. The frequency integral is taken over a spectral window of width $2\delta$ that is chosen to ensure $\Delta\Omega>2\delta\gg\Gamma$, after first subtracting the flat background imprecision. Using equation\,\ref{bp},  we then evaluate the ratio of the vacuum optomechanical coupling rates $g_i$ of the two modes.  In figure \ref{g0s}.b we plot the ratio $g_{1}/g_{2}$ of the optomechanical coupling rates of the two modes. It clearly shows a reduction of the ratio for longer notch lengths, indicating that the coupling rates become more dissimilar as the modes are better hybridized. We compare the experimental data to simulations where we mimic different amounts of disorder, by varying the relative material density of one of the nanobeam slices with respect to the other half. The simulated optomechanical coupling rates are evaluated by extracting the optical cavity resonance shift induced by a known displacement of each eigenmode. A disorder of $2\%$ is found to be a typical value for fabricated devices. For comparison we also show a best case scenario of $0.2\%$. The experimental data fall within the two curves, indicating that structures of intermediate notch lengths exhibit smaller disorder. The reason for the seeming deviation from the trend at notch length of 10 $\mu$m is not fully clear: it could be of statistical origin, but we also note that a strong alteration of dissipation observed in this design may point to other mechanisms that could be at play.  Moreover, long notches design may be more susceptible to other types of phenomena that have not been taken into account such as deformations, buckling, and coupling to other mechanical modes.


To construct the figure of merit for CW measurements we also measure the frequency term ratio of equation\,\ref{fig_merit_CW}, depicted in figure \ref{g0s}.a. To compare this result to a simulated prediction, we take into account the discrepancy between simulated absolute frequencies and the experimentally measured values. We therefore rescale the predicted frequencies by a global factor such that the predicted and measured frequencies match for $0~\mu$m notch length.. Note that for zero extension, the separation between the two modes is very small ($\sim 5$ kHz ) making the denominator of the factor blow up to a singularly high value, making numerical predictions meaningless. Again, the experimental observations follow the trend predicted by the simulations.

Finally we plot the figure of merit for CW measurements in figure \ref{g0s}.c. Increasing the length of the notch has the effect to dramatically decrease the value of the spectral density of mode 1 at the frequency of mode 2. The method therefore shows to be successful in reducing the spurious mode noise contribution. We note that the noise suppression with the longest notch geometry (10 $\mu$m) is less successful because of the amount of disorder of that specific device. The stochastic nature of the occurrence of disorder means that potentially more data points would be needed to draw definitive conclusions on the distribution of occurrences of a given disorder amount and dependencies that arises from the geometry. Structures with intermediate notch lengths suffer less from disorder. Moreover, we note that they are easier to fabricate than longer notches, whose length could play a role for other types of phenomena that have not been taken into account such as deformations, buckling, and coupling to other mechanical modes.

\section{Noise in optomechanical pulsed measurements induced by spurious modes}
\label{sec:pulsed}

In this section, we evaluate the usefulness of the design approach towards a different type of optomechanical experiments, i.e. pulsed measurement. In pulsed optomechanical measurements, the mechanical resonator is probed by light pulses that are shorter than the mechanical oscillation period \cite{Vanner2011PulsedOptomechanics.}. The essence of those backaction-evading measurements is the ability to predict the outcome of a measurement of a single mechanical mode quadrature, given a previous measurement of the same quadrature. This is an example of (pure) state preparation through measurement \cite{Vanner2011PulsedOptomechanics.}. The prediction accuracy (i.e., the purity of the prepared quadrature) $\Delta x$ is given by the difference between the outcome of a verification position measurement $x(t)$ compared with a conditioning position measurement $x(0)$ done beforehand, $\Delta x = x(t)-x(0)$. For a single mechanical mode, best prediction accuracy is achieved whenever the delay between the conditioning and the verification measurement is a multiple of $\pi/\omega_2$, without fundamental precision limit \cite{Vanner2011PulsedOptomechanics.}. However, the difference between the two mode frequencies implies a mechanical beating that in turn implies an extra inaccuracy, except at rare timings that are a common multiple of $\pi/\omega_1$ and $\pi/\omega_2$. We quantify the mechanical decoherence added by a second mechanical mode by considering the expected variance of $\Delta x$ for two mechanical modes at any delay which is extracted through measurement of the cavity frequency $\omega_c=\bar\omega_c+\sum\delta\omega_i(x_i)$, where $\delta\omega_i$ is the cavity frequency shift due to the mode $i$ and $\bar\omega_c$ the cavity frequency for all $x_i =0$. We write the displacement-induced shifts $\delta\omega $ in terms of the mechanical quadratures $X_{m,i} , Y_{m,i}$ of the two modes ($i =1, 2$):
\begin{equation}
\delta\omega= \sum_i G_i (X_{m,i}\cos\Omega_i t+P_{m,i}\sin\Omega_{i} t)
,\end{equation}
Ideally, if only mode 2 couples to the cavity, its quadratures can be determined from measurement of the cavity frequency through $\delta\omega/G_2$ . The thermal fluctuations of other modes evolving at different frequencies adds uncertainty to such estimations. For thermal motion, the variance of the individual quadratures is given by $\left<X_{m,i}^2\right>\simeq x_\mathrm{zpf}^2 k_B T/\hbar \Omega_i=x_\mathrm{zpf}^2 n_{\mathrm{th},i}$, in the limit where $n_\mathrm{th}\gg 1$. Consequently
the variance due to two contributions is the sum of their variances. At a generic
time $\tau$, the position of mode $i$ relative to its position at $t =0$ can be written as
\begin{eqnarray}
\Delta x_{\tau,i} &= x_i(\tau)-x_i(0)\\
&= \sqrt{2} x_{\mathrm{\mathrm{zpf}},i} \left[X_{m,i}(\cos\Omega_i \tau -1) + P_{m,i} \sin\Omega_i\tau\right]
.\nonumber\end{eqnarray}

\begin{figure}
\centering
\includegraphics[width=0.8\columnwidth]{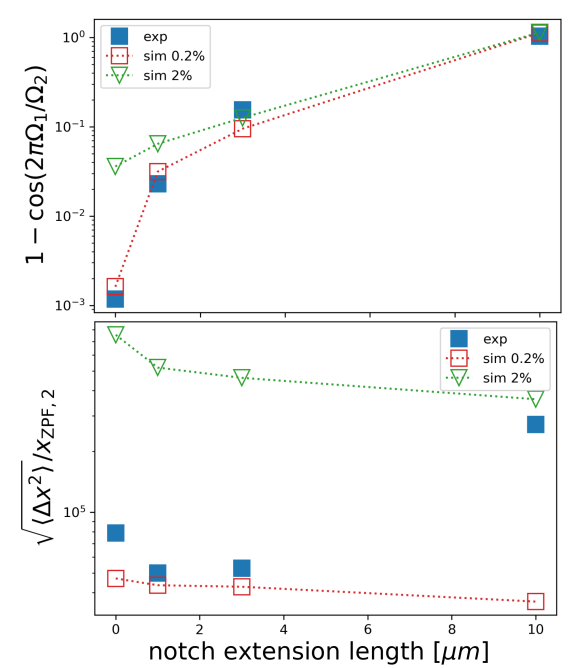}
\caption{(a) Beating factor appearing in equation \ref{fig_merit_nor2}. (b) Figure of merit of spectral purification of a pulsed measurement scenario. Two simulations curves are shown: for disorder equal to $\mathrm f/f=0.2\%$ empty red squares dashed line, and for disorder equal to $\mathrm d f/f=2\%$ empty green triangles dashed line. Blue squares is experimental data.}
\label{PulsedPlot}
\end{figure}
By definition $X_{m,i} , Y_{m,i}$ are independent variables, which means that their cross-correlations vanish. Then the variance of the total position difference writes
\begin{eqnarray}
\left<\Delta x_{\tau,i}^2\right> &= 2 \left<X_{m,i}^2(\cos\Omega_i \tau -1)^2 + Y_{m,i}^2 (\sin\Omega_i\tau)^2\right>\\
&= 4 x_{\mathrm{zpf},i}^2 n_{\mathrm{th},i} (1-\cos\Omega_i\tau)
.\end{eqnarray}
The second line is valid for thermal states, where $\left<X_{m,i}^2\right>=\left<Y_{m,i}^2\right>$ due to the equipartition theorem. 
This means that the variance on the difference $\Delta\omega_c$ between two measurements of the cavity frequency taken at time $\tau$ apart reads 
\begin{equation}\label{pul_noise}
    \left<\Delta\omega_\tau^2\right>=\sum_i 4 g_{i}^2 n_{\mathrm{th},i}\left(1-\cos(\Omega_i\tau)\right)
\end{equation}
in absence of decoherence during $\tau$.
From equation \ref{pul_noise} we see that if the mechanical modes have comparable frequency and optomechanical coupling, they have nearly identical contributions to the variance when $\tau=T_i/4$ or $\tau=T_i/2$, since for delays around those values, $1-\cos\Omega_i\tau = 1$ or $1-\cos\Omega_i\tau = 2$, respectively. Importantly, for $\tau=T_i$, where for a
single mode the difference vanishes, the other mode $j$ still contributes to the variance of the difference
\begin{equation}
\left<\Delta \omega_{T_i}^2\right>=4 g_{j}^2 n_{\mathrm{th},j} \left(1-\cos\frac{2\pi T_i}{T_j}\right)
\label{fig_merit}
.\end{equation}
In the case of only two modes, we have the figure of merit
\begin{equation}
\frac{\left<\Delta x^2_{T_2,1}\right>}{x_{\mathrm{zpf},2}^2}=\frac{\left<\Delta \omega_{T_2}^2\right>}{g_{2}^2}=4\frac{g_{1}^2 }{g_{2}^2} n_{\mathrm{th},1} \left(1-\cos\frac{2\pi \Omega_1}{\Omega_2}\right)
\label{fig_merit_nor2}
,\end{equation}
where we again compare the uncertainty added by the spurious mode to the cavity frequency shift $g_{2}$ of a quantum-level displacement of the mode of interest.
Equation \ref{fig_merit_nor2} scales again with the ratio of vacuum coupling rates of both modes, with the mean occupation number of the unwanted mode, and with the beating factor $\left(1-\cos\frac{2\pi \Omega_1}{\Omega_2}\right)$ that measures the relative dephasing of the two modes, i.e. it is equal to zero when the two modes are exactly degenerate.

We report the dephasing factor of \ref{fig_merit_nor2} for pulsed measurements in figure\,\ref{PulsedPlot}.a. Compared to the simulation, the monotonic increasing trend is reproduced, even though in absolute value the experimental points are smaller since the spread in frequencies increases slower in the experiment than in the simulation. In figure \ref{PulsedPlot}b we show the full expression of the figure of merit. Our method is modestly successful in reducing the spurious mode contribution for relatively small values of the length of the notch extension. For longer beams, we pay the penalty that the occupation number of the common mode $n_\mathrm{th,1}=k_B T/\hbar \Omega_1$ increases since now the frequency of the common mode is much smaller. Due to the larger frequency difference the rate at which the two modes dephase also increases the degree to which the first mode adds uncertainty to the prediction of a quadrature of the second mode after one full period. We remark that after some longer time, close to $2\pi/\Delta\Omega$, the mechanical modes regain the phase difference that they exhibited at $t=0$. The increased frequency difference $\Delta\Omega$ in the notched design will reduce the time at which this occurs, potentially to within thermal decoherence times at practical temperature. This could, in some cases, make the design approach more effective also for pulsed measurement. 

\section{Light-mediated mechanical coupling}
\label{sec:coupledmode}

\begin{figure*}
\centering
\includegraphics[width=\textwidth]{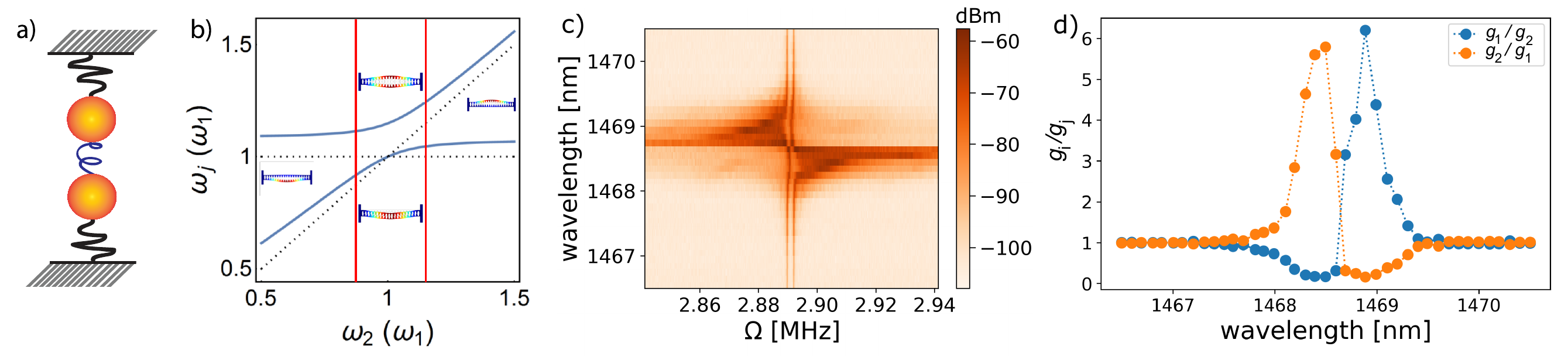}
\caption{a) Illustration of two coupled oscillator. Each oscillators is represented by as a mass attached to a spring. In turn, the two oscillators share a spring that couples them to each other.  b) In the presence of coupling, independent resonant modes $\omega_{1,2}$ (dashed lines) are hybridized into coupled modes (blue solid lines) with frequencies $\omega_j$, which is plotted here as one of the bare frequencies $\omega_2$ is varied. In the left and in the right side of the plot the hybrid modes resemble the bare modes, while in the center of the plot the hybrid modes clearly show a distinct behaviour. The shape of the normal modes in these regimes are shown for the sliced nanobeams. (c) Optical spring effect. Spectra of mechanical motion expressed in dBm (color scale), as a function of laser wavelength with incident power $P_\mathrm{in}=34.5\mu$W. (c) Ratio between single photon-phonon coupling efficiencies versus laser wavelength.}
\label{fig:coupled_modes}
\end{figure*}

We introduced so far the general design principle that reduces symmetry breaking due to imperfections from perturbation theory. A different but related perspective on this approach is gained from the viewpoint of coupled mode theory. We can view the mechanical system depicted in figure\,\ref{fig:disorder} as hosting two mechanical modes, i.e. the lowest-order flexural vibration of each individual string, which are weakly coupled through the clamping support. The modes hybridize only if the coupling strength overcomes the intrinsic frequency detuning of the two strings, which is related to fabrication imperfection. The frequency splitting for degenerate strings $\Delta\Omega$ is equal to twice the rate $J$ at which the two strings are coupled. Thus, if one increases the mechanical coupling rate, the hybridization into even and odd modes becomes more resilient to disorder.

The mechanical designs we presented thus far can only be described to a limited extent from this specific viewpoint, as for larger notch lengths the new eigenmodes can no longer be described as even and odd superpositions of the individual string modes of the structure without a notch. However, we can imagine a different coupling mechanisms, in which we can describe them in this way. In this section, we explore the use of mechanical coupling of the individual strings through radiation pressure, which can lead to normal mode splitting and associated hybridization as demonstrated by Shkarin et al.\cite{Shkarin2014OpticallyModes}. Due to the orthogonality of the hybrid modes and the spatial dependence of the optomechanical coupling, such induced coupling can render one of the two new modes optomechanically dark. We demonstrate that principle below for our structures, and evaluate the merits of this optically-induced coupling for single-mode measurement.

In  figure\,\ref{fig:coupled_modes}b, we show a generic prediction of \textit{strong coupling} for two normal modes at frequencies $\omega_{1,2}$, characterized by the Rabi frequency splitting for $\omega_1 \simeq \omega_2$, that quantifies the coupling rate between the modes. In the presence of strong coupling, coupled modes are generated, which converge to the original modes when the frequency difference $\omega_2 - \omega_1$ is large, far away from the strong coupling region which is larger the larger the Rabi splitting between the coupled modes.  The coupled modes represent another set of orthogonal modes in which phase relations between the individual bare modes are established. For the nanobeams, the two uncoupled modes, characterized by the independent vibration of the individual nanobeam slices, are transformed by the coupling into an even and an odd normal mode.

To understand the mechanism, let us consider the interaction of this system of two mechanical modes with the optical field in an optomechanical system. When the laser frequency is within the optical cavity linewidth the radiation pressure force exerted by the driving light causes an optical spring shift that causes the mechanical frequency to decrease (increase) for red (blue) detuning. We note that since we are in the regime of nonlinear transduction, we typically observe an associated broadening of the mechanical resonances which become asymmetric at room temperature because of the thermal fluctuation of the spring shift magnitude \cite{Leijssen2017NonlinearFluctuations}. For cavities in this regime, the optical spring effect cannot be estimated with a model based on linearized equations of motion as the magnitude of the spring shift varies with the particular amplitude of oscillation at a given time. The behaviour in this case can be accounted for by considering the thermal distribution of the mechanical oscillation amplitude. This is done by calculating an effective spring constant from the distribution of results obtained by modeling the spring shift using the first Fourier coefficient of the radiation pressure force \cite{Leijssen2017NonlinearFluctuations}. The light field that is modulated by the motion of one of the two mechanical modes will in turn drive the other mechanical mode. If the spring shift is such that it tunes the modes to spectrally overlap, the two modes can become degenerate. The resulting optically mediated coupling can give rise to the creation of mechanical hybrid modes with opposite symmetry. The magnitude of the induced coupling rate is 
\begin{equation}
J_{opt} = 2 g_{1}g_{2}n_c/\kappa
,\end{equation}
where $n_c$ is the number of intracavity photons. The effect is closely connected to the spring effect for a single mechanical resonator \cite{Aspelmeyer2014CavityOptomechanics} .

In figure \ref{fig:coupled_modes}.c we show optomechanical spectra as we scan the laser wavelength across the cavity resonance. The spectra depict the behaviour for two different regimes of power for a structure for which the separation between the frequencies of oscillation of the individual beam slices is only $\Delta\Omega=\Omega_2-\Omega_1 \simeq 2.19$ kHz. The individual, uncoupled, behaviour of the two modes is evident far from optical resonance. Around optical resonance, in the region where the optical spring shift  $\delta \Omega$ is larger than the frequency separation $\delta \Omega \gg \Delta\Omega$, we observe strong coupling between the two modes. The coupling is mediated by the light since the spring shift is caused in the first place by the optomechanical interaction. In particular, for each value of the detuning we observe that only one of the modes is bright, i.e. strongly coupled to the cavity, while the other is relatively suppressed. Two main observations support this conclusion: on the one hand we observe a strong spring shift for only one of the two modes (the bright one), on the other hand, we observe an asymmetric broadening of the linewidth for the same mode \cite{Shkarin2014OpticallyModes}.
To address how the optically induced coupling modifies the response of the system we look at the single photon-phonon coupling efficiencies for the two modes as a function of laser wavelength  (figure\,\ref{fig:coupled_modes}.c). In order to evaluate the $g$ ratio we use equation\,\ref{bp} and the data of figure\,\ref{fig:coupled_modes}.b
\begin{equation}
\frac{g_{i}}{g_{j}}=\sqrt{\frac{BP_i \Omega_i}{BP_j \Omega_j}}
,\end{equation}
For the band power $B P_i$ we take the area underneath the mechanical peak of mode $i$ at a specific laser wavelength, and for the frequency $\Omega_i$ we take the frequency corresponding to the maximum value for the $P_{ESA}$ . From figure\,\ref{fig:coupled_modes}, we observe that the coupling of the two mechanical modes to the light field becomes strongly asymmetric as the light mediated coupling forces one of the coupled modes to become optomechanically dark.

In order to compare this scenario to the results for the design presented previously we evaluate the figures of merit. The ratio $g_{2} /g_{1}$ is less than or equal to 6. For continuous measurements this corresponds to having a noise contribution from mode 1 of $S_{\omega\omega}^\mathrm{th,1}/S_{\omega\omega}^\mathrm{zpf,2}/\Gamma_1\Gamma_2\geq 3\times10^{20}$\,$\mathrm{Hz}^{-2}$ which is better by a factor of $\sim 10^4$ with respect to the off resonance $\Delta\neq0$ or low power cases. Here we disregard enhancement of the apparent mode linewidths, as this effect would be absent at low temperatures where the thermal fluctuations are reduced \cite{Leijssen2017NonlinearFluctuations}. For pulsed measurements the noise contribution is $\left<\Delta\omega_{T_2}^2\right>/g_{2}^2\geq3.8\times10^3$ which is a factor of $\sim2.5$ better than for off-resonance and low power configurations. We remark that the results presented in this section correspond to a device for which the disorder induced by the fabrication is exceptionally low. In this circumstance, there is a clear benefit for pulsing experiments since the figure of merit goes to \textit{zero} for degenerate modes.

We conclude that by using a second laser that serves as a coupling laser, the noise contribution for pulsed measurements can be in principle brought to a value that is lower than for the notched structures. We do note that for quantum-limited operation, the influence of that coupling laser in terms of radiation pressure shot noise fluctuations and heating through absorption should be considered carefully.

\section{Conclusion}

We reported a mechanical design strategy that produces mechanical mode symmetries robust against imperfections. Our method is inspired by perturbation theory, which suggests that mechanical spectral control can be used to minimize the symmetry breaking that is due to fabrication disorder. The goal of this strategy is to minimize the detrimental effects due to the presence of a spurious mode in optomechanical measurements. We defined figures of merit for continuous  and pulsed measurement schemes, that take into account both the values of the optomechanical coupling strength of the different modes and their frequencies. We implemented the reported strategy in practical realizations of sliced nanobeam cavities, through mechanical notched design and through coupling via radiation pressure. By experimentally evaluating the figures of merit, we conclude that the strategy is expected to greatly mitigate spurious mode noise in continuous measurement, and moderately for pulsed ones. This design strategy can be adapted to generic optomechanical systems.

\ack
This work is part of the research program of the Netherlands Organisation for Scientific Research (NWO). The authors acknowledge support from NWO Vidi, Projectruimte, and Vrij Programma (Grant No. 680.92.18.04) Grants, the European Research Council (ERC starting Grant No. 759644-TOPP), and the European Union's Horizon 2020 research and innovation program under Grant Agreement No. 732894 (FET Proactive HOT).

\appendix

\section{\textbf{Noise in continuous measurements induced by a spurious mode}}
\label{ContinuousFOM}

The spectral density of optical frequency fluctuations $S_{\omega\omega}(\Omega)$ is related to the displacement spectral densities $S_{x_ix_i}(\Omega)$ of one or more mechanical degrees of freedom labeled by index $i$. In the case of two uncorrelated mechanical modes ($i=1,2$)
\begin{equation}
S_{\omega\omega}(\Omega) = G_1^2 S_{x_1x_1}(\Omega) + G_2^2 S_{x_2x_2}(\Omega)
.\end{equation}

The response of resonator $i$ to a driving force $F_i$ is described by the susceptibility $\chi_i(\Omega)$, such that
$x_i(\Omega) = \chi_i (\Omega) F_i(\Omega)$
where
\begin{equation} \label{suscept}
\chi_i(\Omega) = \frac{1}{m_i(\Omega_i^2- \Omega^2) - i m_i \Gamma_i\Omega}
,\end{equation}
with $m_i$ the mass of the mechanical oscillator and $\Gamma_i$ the linewidth. 
This means that the power spectral densities of force and displacement are related through
$ S_{x_i x_i}(\Omega) =  \left|\chi_i(\Omega)\right|^2 S_{F_iF_i} $. This allows prediction of the displacement spectral density due to an external fluctuating force described by $S_{F_iF_i}$.
The resulting optical frequency spectral density is thus
\begin{equation}
S_{\omega\omega}(\Omega) = G_1^2\left|\chi_1(\Omega)\right|^2 S_{F_1F_1} + G_2^2 \left|\chi_2(\Omega)\right|^2 S_{F_2F_2}
,\end{equation}
where numbers in subscripts refer again to the two mechanical modes whose contributions can be described independently.

We consider here the frequency noise due to the thermally driven mechanical mode $i=1$, caused by the thermal Langevin force with fluctuations $S_{F_1F_1} = 2 \Gamma_1 m_1 k_B T$, where $k_B$ is the Boltzmann constant and $T$ the bath temperature. The cavity frequency noise it induces has spectral density
\begin{equation}
S_{\omega \omega}^\mathrm{th,1} (\Omega)=G_1^2\left|\chi_1(\Omega) \right|^2 S_{F_1F_1} 
\end{equation}
$$= \frac{4\Gamma_1 k_B T g_\mathrm{1}^2}{\hbar}  \frac{\Omega_1}{(\Omega^2-\Omega_1^2)^2+\Gamma_1^2\Omega^2} ,$$
where we used (\ref{suscept}) and $g_{1}^2=G_1^2\hbar/(2m_1\Omega_1)$.

To define a figure of merit, we consider the noise $S^\mathrm{th,1}_{\omega\omega}(\Omega_2)$ due to the spurious mode as a background to the measurement of the mode of interest ($i=2$) at frequency $\Omega_2$. We quantify this background in units of the spectral density $S^\mathrm{zpf,2}_{\omega\omega}=2g_{2}^2/\Gamma_2$ of the cavity fluctuations related to the zero-point fluctuations of mode 2. Normalizing the spurious noise to the level of mechanical quantum fluctuations has broad relevance to quantum measurement, e.g. if the goal is to characterize the quantum state of resonator 2 or if one wants to cool it to the ground state through feeding back a measurement record. The figure of merit thus reads
\begin{equation}
\frac{S_{\omega\omega}^\mathrm{th,1}}{S_{\omega\omega}^\mathrm{zpf,2}}(\Omega_2)=2\frac{k_B T}{\hbar} \Gamma_1\Gamma_2 \frac{g_{1}^2}{g_{2}^2} \frac{\Omega_1}{(\Omega_1^2-\Omega_2^2)^2+\Gamma_1^2\Omega_2^2} 
.\end{equation}

Defining $\Delta\Omega\equiv\Omega_2-\Omega_1$ and assuming that the mechanical modes are spaced by more than their linewidths ($\Gamma_1\ll\{\Omega_1,\Delta\Omega\}$), the figure of merit can be approximated by
\begin{equation}
\frac{S_{\omega\omega}^\mathrm{th,1}}{S_{\omega\omega}^\mathrm{zpf,2}}=2\frac{k_B T}{\hbar} \Gamma_1\Gamma_2 \frac{g_{1}^2}{g_{2}^2}\frac{\Omega_1}{\Delta\Omega^2(\Delta\Omega+2\Omega_1)^2}.
\end{equation}

\section*{References}
\bibliography{bibliography.bib}

\begin{thebibliography}{10}
\bibitem{Note}
One notes that the figure of merit derived assumes that $\omega_2>\omega_1$. a
  slightly different figure of merit should be used for the other regime
  because of the fact that the mechanical susceptibility is not symmetric
  around resonance for large $|\delta\omega|$. such a fom could be defined in a
  similar fashion following the calculation derived in appendix.
\end{thebibliography}

\end{document}